\documentclass[12pt,a4paper]{article}
\usepackage[english]{babel}
\usepackage{graphics}
\usepackage{amssymb}
\newfont{\cc}{cmbsy10 scaled 1000}
\newcommand{\A}{{\bf a}}
\newcommand{\C}{{\mathbb C}}
\newcommand{\ie}{{\it i.e.} }

\newcommand{\nbd}{neighbourhood }

\newcommand{\LL}{{\cal L}}
\newcommand{\Ln}{{\mathbb L}(n)}

\newcommand{\mm}{{\bf m}}
\newcommand{\nn}{{\bf n}}
\newcommand{\N}{{\mathbb N}}
\newcommand{\R}{{\mathbb R}}
\newcommand{\Ss}{{\mathbb S}}
\newcommand{\T}{{\mathbb T}}
\newcommand{\Z}{{\mathbb Z}}
\newcommand{\pdint} {\rule{5pt}{0.5pt}  \rule{0.5pt}{6pt}\;}
\newcommand{\ptir}{\discretionary{.}{}{.\kern.35em---\kern.7em}}
\newcommand{\cqfd}{\hfill\rule{5pt}{5pt}}

\newtheorem{theo}{Theorem}

\newtheorem{lemma}{Lemma}[section]
\newtheorem{prop}{Proposition}[section]

\newtheorem{cor}{Corollary}[section]

\begin{document}
\begin{titlepage}
\title{Bohr-Sommerfeld conditions for several commuting Hamiltonians}
%\title{Joint spectrum and periodicity of the Hamiltonian flow in semiclassical calculus}
%\title{Localization of the joint spectrum of several commuting h-pseudodifferential operators
%with a periodic flow on a given energy level }
\author{}
\date{\today}
\maketitle
\thispagestyle{empty} 

\begin{center}
{\bf Colette ANN\'E }  \\
UMR 6629 de Math\'ematiques, BP 92208 \\ 
Universit\'{e} de NANTES \\ 
Facult\'{e} des Sciences et des Techniques \\
44322 Nantes-Cedex 03, France\\
\end{center}
\begin{center}
{\bf Anne-Marie CHARBONNEL} \\
UMR 6629 de Math\'ematiques, BP 92208 \\ 
Universit\'{e} de NANTES \\ 
Facult\'{e} des Sciences et des Techniques \\
44322 Nantes-Cedex 03, France\\
\end{center}\vskip1cm
\begin{abstract}
The goal of this paper is to find the quantization conditions of Bohr-Sommerfeld of several 
quantum Hamiltonians ${ Q_1(h), ...,Q_k(h)}$ acting on $ { {\R}^n}$, 
depending on a small parameter h, and which commute to each other. That is we determine, around a regular 
energy level $E_0\in \R^k$ the principal term of the asymptotics in $h$ of  eigenvalues $\lambda_j(h),\, 
1\leq j\leq k$  of the operators $Q_j(h)$ that are associated to a common eigenfunction. Thus
we localize the so-called joint spectrum of the operators.

Under the assumption that the classical Hamiltonian flow of the joint principal 
symbol $q_0$ is {\it periodic with constant periods on the one energy level} $q_0^{-1}(E_0)$, 
we prove that the part of the joint spectrum lying in a small neighbourhood of $E_0$ is localized 
near a lattice of size $h$ determined in terms of actions and Maslov indices. The multiplicity 
of the spectrum is also determined.

\end{abstract}
\vfill
MSClass : 81Q20, 35S30, 47G30.
\end{titlepage}
\newpage
\thispagestyle{empty} 
\tableofcontents
\vfill
\newpage
\section{Introduction}

The joint spectrum of several commuting operators arises naturally if suitable symmetries are present. 
Let us consider for instance the Schr\"odinger operator, acting on $\R^3 $  
 $$ A_1 (h) \ = \ -h^2 \Delta \ + \ V(x) \ ,  $$
and suppose that the potential is  realvalued, smooth, and spherically symmetric. 
 Then $ A_1 (h) $ commutes with the kinetic momentum  
 $$ A_2 (h) \ = \ -ih \ (x_2 \partial _{x_3} \ - \ x_3 \partial _{x_2} )  . $$
The bounded states of $A_1 (h)$ are common eigenfunctions of  $A_1 (h)$ and  $A_2 (h)$, 
and if we consider a third operator 
$$ A_3 (h) \ = \ -h^2 \ \lbrace (x_2 \partial _{x_3} \ - \ x_3 \partial _{x_2} )^2 \ + \ (x_3 \partial _{x_1} \ - \ x_1 \partial _{x_3} )^2 \ + \ (x_1 \partial _{x_2} \ - \ x_2 \partial _{x_1} )^2 \ \rbrace $$
which commutes with the other two operators, these bounded states are common eigenfunctions of $A_1 (h)$, $A_2 (h)$ and $A_3 (h)$, and their joint spectrum consists of the associated 3-uplets of eigenvalues,  
$$ \lambda (h) \ = \ ( \lambda_1 (h), \lambda_2 (h), \lambda_3 (h)) \in \R^3 .$$  
They have a multiplicity equal to 1, the bounded states are said to be completely separated. 
\par \noindent
In fact this example is integrable (the number of operators equals the dimension) and by the Liouville 
theorem 
the hamiltonian flow is periodic, (see \cite{CH3} for the explicite calculation of the joint flow for the 
harmonic oscillator). This example can naturally be extended to all dimensions.
Few new examples of integrable systems on manifolds, inspired by the classical case can be found in
the works of J.A. Toth \cite{T1} and \cite{T2}.

On the other hand  with a potential of the form $V(x,y)=f(\|x\|^2)+g(\|y\|^2)$ with $x\in\R^k$ 
and $y\in\R^{n-k}$ we can construct examples with less symmetries, and which are nomore integrable.
%%%%%%%%%%%%%%%%%%%%%%%% pseudo  %%%%%%%%%%%%%%%%%%%%%%%%%%%%%%%%%
\subsection{Pseudodifferential calculus} 
More generally we consider k quantum hamiltonian ${ Q_1(h), ...,Q_k(h)}$ acting on $ { \R^n}$, 
depending on a small 
parameter h, and which commute to each other.
The $Q_1(h), ...,Q_k(h) $ are supposed to be $h$-admissible pseudodifferential operators, as 
introduced by B. Helffer and D. Robert in \cite{HR1} and \cite{HR2} (see \cite{robert} for the 
general theory). 
\par \noindent Let $p$  be a weight function on $\R^{2n}=T^*\R^n$, \ie a continuous function 
from $\R^{2n}$ into $\R^{+}$ which  satisfies the following property : 
\par \noindent there exist constants $C > 0 , \ m > 0$ such that 
$$
\forall(x, \xi),(x', \xi') \in \R^{2n} \ p(x, \xi) \ \leq \ C \ p(x', \xi') \  ( 1 + 
\vert x-x'\vert ^2  + \vert \xi-\xi' \vert ^2)^m \ \ .
$$
We denote by $ S_{p} $ the space of all the functions $ a \in C^{\infty}(\R^{2n}) \ $ such that 
for all $  (\alpha , \beta) \in { \N }^{2n} $ there exists $ C_{\alpha , \beta } > 0 $ satisfying
\begin{equation} 
\vert \partial _{x}^{ \alpha}\partial _{\xi}^{ \beta} a (x,\xi)  \vert \ \leq \ C_{\alpha , \beta } 
\ p (x,\xi) \quad  \forall (x,\xi) \in \R^{2n} .
\end{equation}
To such a symbol $a$ we associate a $h$-pseudodifferential operator $ { \cal A} (h)  =   Op_h^w (a) $
acting on functions $\psi \in {\cal S}(\R^n),$ the Schwartz space of smooth functions rapidly
decreasing at infinity together with their derivatives, defined by the Weyl quantization~:
$$
 { \cal A} (h) \psi(x) \ = \ (2\pi h)^{-n}\int\int e^{i h^{-1}\langle x-y , \xi\rangle} \,
a ({{x+y}\over 2 } ,\xi) \, \psi (y) \, dy \, d\xi  \ .
$$

More generally we may assume that the symbol $a$ depends on $h$~: if there exists a sequence 
$\{ a_{i}\}_{i \in \N}  \in  S_p $ such that  
\begin{equation} 
a (h,x,\xi) \ \sim \ \sum_{i\geq 0} h^{i} a_{i}(x,\xi)
\end{equation} 
where $\sim$ means that for all $N \in \N$ the difference 
$$
r_{N+1}=h^{-(N+1)} \ \big( \ a(h,x,\xi) \ - \ \sum_{i = 0}^N h^{i} a_{i}(x,\xi) \ \big)
$$
defines a family ${\cal R}_{N+1}(h)=Op_h^w (r_{N+1}),\, h\in]0,h_0]$ of {\em bounded} operators
which is  bounded (uniformly in $h$) in $\LL(L^2(\R^n))$; then we say that the operator $ { \cal A} (h)$
is {\em $h$-admissible.}

We suppose that we have $  Q_j (h)  =   Op_h^w (q_j) $ are $h$-admissible, so there exists a sequence 
$\{ q_{ij}\}_{i \in \N}  \in  S_p $ such that  
\begin{equation} 
q_j (h,x,\xi) \ \sim \ \sum_{i\geq 0} h^{i} q_{ij}(x,\xi).
\end{equation} 
\par \noindent Then $ Q_j (h) $ belongs to $ {\cal L } \big( {\cal S} (\R^n),{\cal S} (\R^n) \big) $ 
for each $j = 1, \ldots ,k$. 

We also assume that each operator $Q_j(h)$ admits a {\em self-adjoint extension} to 
$L^2 (\R^n)$. It is the case if the principal symbol $q_{0j}$  is itself a weight and if 
$q_{ij}\in  S_{q_{0j}}$ for all $i$, and more generally if $q_{0j}$ is bounded from below \cite{robert}.

In our case, we always can reduce to this situation  by means of the 
functional calculus, if the weight is $p=\|q_0\|^2+1$, and if the operator $\sum_{j=1}^k {Q_j}^2(h)$
is $h$-admissible~: let $f$ be a $C^\infty$ function on $\R^+$ such that $f(x)=x$ for $x\in[0,c]$, 
(with $0<c<1$) $f$ is increasing and asymptotic to 1 at infinity and define $ g(x)= f(x)/x$ a 
$C^\infty$ function on $\R^+$, equal to 1 on the interval $[0,c]$. Then, for $E_0\in\R^k$, the new 
operators defined vectorially by
\[P(h)=g(\|Q(h)-E_0\|)(Q(h)-E_0)
\]
are $h$-admissible pseudodifferential operators (remark that the operators $g(\|Q(h)-E_0\|)$ are bounded) 
with bounded symbols, so every principal symbol is a weight bounded from below. Moreover the $P_j(h)$
commute to each other if the $Q_j(h)$ have this property~; furthermore, on the domain
$\{(x,\xi), \|q(h)(x,\xi)-E_0\|\leq c\}$ one has $p(h)=q(h)-E_0,$ so $p(h)$ is proper if $q(h)$ is.
%%%%%%%%%%%%%%%%%%%%%%%%   flow   %%%%%%%%%%%%%%%%%%%%%%%%%%%%
\subsection{Commutativity and Hamiltonian flow} 
\par \noindent We suppose that the operators $Q_j(h)$ commute to each other 
\begin{equation}
\lbrack Q_i(h) , Q_j(h) \rbrack =  0 \ {\rm on} \ \ {\cal S} (\R^n) \ \forall \ i , j = 1, 
\ldots , k \ , \ \forall \ h > 0 \ \ \ .
\label{eq:1.3}
\end{equation}
\par \noindent We denote by $q_0 = (q_{01} , q_{02}, \ldots , q_{0k}) \ : \ T^*\R^n \ \rightarrow \ 
\R^k$ the joint principal symbol of the 
operator $Q (h) = (Q_1(h) , Q_2 (h) , \ldots , Q_k(h))$, and by $q_1 = (q_{11} , q_{12}, \ldots , 
q_{1k})$ its sub-principal symbol. \par \noindent 
As a consequence of the commutativity, one has $\{  q_{0i} , q_{0j}\} = 0$, and the associated 
symplectic or classical flows $\Phi_j^t$ also commute. 

The $\Phi_j^t$ are defined as follows~: with the canonical symplectic form $\omega$ of $T^*\R^n$ : 
$\omega = \sum_{j=1}^n d \xi_j \wedge dx_j $ one can define the Hamiltonian field $H_j$ of $q_{0j}$ by~:
\begin{equation}
d q_{0j} \ = -H_j \ \ \pdint \ \ \omega
\label{eq:1.5}
\end{equation}
then the $\Phi_j^t$ are the symplectic transformations satisfying~:
\begin{equation}
\Phi_j^0 = Id \ ; \ {{\partial } \over {\partial t }} \Phi_j^t = H_j \circ \Phi_j^t.
\label{eq:1.6}
\end{equation}

\par \noindent We want to look at spectral properties for the operators $Q_j$ in a neighbourhood of a 
value $E_0 \in \R^k$ on which we make the following assumption : 
\vspace*{0.4cm} 
\begin{itemize} 
\item[$(H_1)$] {\it $E_0$ is a regular value of $q_0$ and $q_0$ is proper in a neighbourhood of 
$E_0$ : there exists a compact neighbourhood $K_1$ of $E_0$ in $\R^k$ such that $q_0^{-1} (K_1)$ 
is compact.}
\end{itemize} 
\vspace*{0.4cm} \noindent 
Note that this assumption  is fulfilled for the example above if  
$$
-\infty< \liminf_{ \vert x \vert \to \infty } V(x).
$$ 
Moreover,  the  conditions above imply $ k \leq n $.

The joint spectrum $\Lambda^{Q(h)}$ of commuting operators, as $Q_j(h)$, has been defined in  \cite{CH1}, 
and it has been proved that, under the hypothesis ($H_1$), the part of the joint spectrum contained in any 
compact $K$ included in the interior of $K_1$ consists in finitely many joint eigenvalues of finite 
multiplicity, \ie   
\begin{equation}
\Lambda^{Q(h)} \bigcap K  =  \lbrace (\lambda _q (h) )_{q \in {  I(h) } } ;  \exists \Psi _q^{h} 
\in  L^2(\R^n)  :  Q_j(h)\Psi _q^{h}  =  \lambda _q^{j} (h) \Psi _q^{h} 
\ \forall  j = 1,.,k \rbrace  
\label{eq:1.4}
\end{equation}
where {  I(h) } is finite.

The asymptotics of these eigenvalues can be precised with an assumption of periodicity of the classical
flow. We will make a very weak assumption as follows.
Suppose that $K$ is sufficiently small to be composed of regular values of $q_0$. As a consequence of 
$(H_1)$, for all $E$ belonging to $K$, the set $\Sigma_E = q_0^{-1}(E)$ has the structure of a submanifold 
of $\R^{2n}$ of codimension k, $\Sigma_E$ is compact and invariant under the action of the Hamiltonian flow
which is symplectic.
Moreover the Hamiltonian fields $H_j$ are independant, and  $H_j \ \ \pdint \ \ \omega$ vanishes
in restriction to the tangent plane of $\Sigma_E$, so this  manifold  is co-isotropic. 

\noindent
The Hamiltonian fields $H_j$ commute so  the joint flow defines finally an action $\rho_E$ of $\R^k$ on 
$\Sigma_E$~:
\begin{eqnarray}
\rho_E\ :\qquad\qquad \R^k\times \Sigma_E &\to& \Sigma_E\nonumber\\
\left((t_1,\ldots,t_k),\nu\right) &\mapsto&\Phi_1^{t_1}\circ\ldots\circ \Phi_k^{t_k}(\nu).
\label{eq:1.7}\end{eqnarray}

\noindent We will suppose that the joint flow is periodic with constant periods on {\em one} level
$\Sigma_0 = \Sigma_{E_0}$. More precisely we make the following hypothesis~:
\begin{itemize} 
\item[$(H_2)$] {\it All the points of $\Sigma_0=q_0^{-1}(E_0)$ are periodic under the action of the 
joint flow $\rho_0 = \rho_{E_0}$ with the same lattice of periods.}
\end{itemize}
Let $(e_1, \ldots , e_k)$  be a basis of this lattice ; it is a 
basis of $\R^k$ verifying for all $\nu \in \Sigma_0$ and for all $z = (z_1, \ldots ,z_k) \in \Z^k$ :
$\rho_0 \ \big( 2 \pi \sum_{j=1, \ldots ,k} \ z_j e_j \ , \ \nu \big) \ = \ \nu.$ 

\noindent So $\rho_0$ can be regarded as the action of a torus $\T^k$ on $\Sigma_0$, that we will still 
denote by $\rho_0$. By the last hypothesis {\em this action is free}, \ie without fixed points.
%In the following, for $\gamma \in \T^k$ and $\nu \in \Sigma_0$, we will write $\rho_0 (\gamma , \nu)$
%instead of $\rho_0 (t,\nu)$, where we have denoted by $\gamma$ the class of $t \in \R^k$ in $\T^k$.

\noindent 
{\sl Comparison with the hypothesis of \cite{CH1} and notations\ptir} 
In this previous work it was made an assumption of periodicity on a open set of energy.
More precisely it was supposed that there exists a function $f \in C_0^{\infty} (\R^k , \R^k )$ 
such that the principal symbols of the operators $f\big( {\it Q} (h) \big)$ are periodic in a 
neighbourhood of the energy level. 

\noindent In our case 
let $(\varepsilon_j)_{1 \leq j \leq k}$ be the canonical basis of $\R^k$ and define ${ \bf a} \in 
{\it End }(\R^k)$ by $e_j= {\bf a}(\varepsilon_j)=\Sigma_i \alpha_{ij}\varepsilon_i$. 
\par \noindent Then the new symbol $p_0 = (p_{01},\ldots , p_{0k}) = {\bf a}(q_0)$ satisfies 
$$ dp_{0j}= \Sigma_i \ \alpha_{ij} \ dq_{0i} = - \Sigma_i \ \alpha_{ij} \  H_i \ \pdint \ \omega $$
and the corresponding Hamiltonian flow $\Psi_j$ is $2 \pi$-periodic on the energy level ${\bf a}(E_0)
$ in view of $(H_2)$.

\noindent Consequently the operators  
${\it P} (h) \ = \ {\bf a} \big( {\it Q} (h) \big)$ 
satisfy the pointwise equivalent of the hypothesis of \cite{CH1}. 
\par \noindent We will denote $F={\bf a}(E_0)$ and 
$K_j \ = \ \Sigma_i \ \alpha_{ij} \ H_i
$ the Hamiltonian field of $p_{0j}$.

We can now give our main result : \par 

\begin{theo}\ptir Let $  Q_j (h)  =   Op_h^w (q_j),\, 1\leq  j\leq k $ be k commuting $h$-admissibles 
pseudodifferential operators essentially self adjoint and satisfying the following hypotheseses~:\\
($H_1$) the joint principal symbol $q_0$ is proper in a neighborhood of a regular value $E_0,$\\
($H_2$) the classical flow is periodic with constant period on the energy level 
$\Sigma_0=q_0^{-1}(E_0),$\\
($H_3$) the subprincipal $q_1$ vanishes,\\
($H_4$) the surface $\Sigma_0$ is  connected.

Then the part of the joint spectrum $\Lambda ^{{\it Q} (h)}$ lying in any
$k$-cube $ \displaystyle { \prod_{j=1}^{k}} \ \rbrack E_{0j} -h c_j \ , \ E_{0j} + h c_j \lbrack$ 
centered at $E_0$ is discrete and localized modulo $O(h^2)$ near a lattice 
$$E_0 \ + \  {\bf a}^{-1} \Big( (-\frac{1}{4} \mu_1 h - \frac{\alpha_1}{2\pi} + \Z h)
\oplus\ldots \oplus ( - \frac{1}{4}\mu_k h - \frac{\alpha_k}{2\pi} + \Z h) \Big) ,$$ 
where the  $\mu_j$ are the Maslov indices of the basic 
cycles of the torus acting on $\Sigma_0$, $\alpha_j$ are the action of these cycles and ${\bf a}\in Gl(k)$. 
\end{theo}

\noindent{\sl Comments}\ptir The hypothesis ($H_3$) can be weaked in
\begin{itemize} 
\item[$(H'_3)$] {\it The integral of the subprincipal symbol $q_1$ on a closed trajectory of the joint 
Hamiltonian flow is 
independent of the point on the energy level $\Sigma_0$, it depends only on the period}. 
\end{itemize}
see the Theorem 2, subsection 2.6. below.

When the hypothesis ($H_4$) failes each connected component gives a part of the discrete spectrum.
%%%%%%%%%%%%%%%%%%%%%%%%%%%%%%% plan  %%%%%%%%%%%%%%%%%%%%%%%%%%%%%%%%%%%%%%%

\noindent{\sl About the method\ptir} 
We will look at the operators $P(h)\ = \ {\bf a} \big( {\it Q} (h) \big)$. 
For these ones the Hamiltonian flow of each component $p_{0j}$ of the principal symbols is 
$2\pi$-periodic, we remark that $\mu_j$ the Maslov index of the $2\pi$-periodic 
trajectories of the Hamiltonian flow of $p_{0j}$ is constant on $\Sigma_0$ as an invariant of homotopy ; 
as well the action of such a trajectory, $\alpha_j,$ is constant on $\Sigma_0$. We will denote by $\mu$ 
respectively
$\alpha$ the $k$-vector defined by the Maslov index, respectively the action, of the basic cycles.

In order to localize the eigenvalues  of $P(h)$ lying in the $k$-cube 
$ \displaystyle 
{ \Pi_{j=1}^{k}}  \rbrack F_j \ - \  ch , F_j \ + \  ch \lbrack $ modulo $O(h^2)$ we suppose for the moment 
that the subprincipal symbols are null and look at the following operator 
$
\zeta ({{P(h)-F} \over {h}}) \ \theta (P(h))  , 
$
where $\hat\zeta$ and $\theta$ belong to $C_0^{\infty}(\R^k)$  and 
$\theta(\lambda) = 1$ for $\lambda$ in a smaller neighbourhood of $F$ and compare it with 
$\zeta ({{P(h)-F} \over {h}})\exp-{{i}\over{h}}<T,P(h)> \ \theta (P(h))$ where 
$T\in 2\pi\Z^k$, the lattice of periods.

\noindent We will do this by  approaching  these operators by Fourier Integral Operator.
We will see that their Lagrangian manifold are identical and their principal symbol differ by a scalar 
which is determined by the Maslov index and the action of the closed trajectories, consequently 
the $L^2$ norm of the difference between $\zeta ({{P(h)-F} \over {h}}) \ \theta (P)  , $ and the operator 
$\zeta ({{P(h)-F} \over {h}}) \exp-{{i}\over{h}}(<T,P(h)>+ \alpha + {{\pi} \over{2}} \mu
h) \ \theta (P)$ ,  is only $O(h)$. 

This method, consisting on espressing the evolution operator $e^{-ih^{-1}<t,P(h)>}$ by means of the 
theory of Fourier Integral Operators was initiated by Duistermaat-Guillemin \cite{DG} and Colin de Verdi\`ere 
\cite{Cdv} for compact manifolds, and Helffer and Robert \cite{HR3} in the semi-classical case. 
We especially tried to make clear the appearance of the Maslov Index.
On the other hand one can find in \cite{ac1}, a construction, by symplectic geometry, of new symbols with 
the strong property of periodicity and which approach the first ones. But the new operators do not commute 
anymore.

This paper is organized as follows : in the second section we will prove the theorem 1, and in the 
third one we will look at the multiplicities of the joint eigenvalues.

%%%%%%%% PROOF %%%%%%%%%%%%%%%%%

\section{Proof of theorem 1}
Let $\zeta$ be a function belonging to ${\cal S} (\R^k)$ with a compactly-supported 
Fourier-transform $\hat{\zeta}$, and $\theta$ belonging to $C_0^{\infty}(\R^k)$ with a compact 
support lying in a neighbourhood of $F$ and verifying
$\theta(\lambda) = 1$ for $\lambda$ in a smaller neighbourhood of $F$. 
We can write 
$$
\zeta ({{F-P(h)} \over {h}}) \ \theta (P(h)) \ = \ {{1} \over {(2 \pi)^k}} \ \int_{\R^k} e^{-ih^{-1}<t,(P(h)-F)>} 
\hat{\zeta}(t) \theta (P(h)) dt \ \
\  . 
$$
Using the functional calculus developed in \cite{CH1}, we  consider $\theta(P(h))$ as a 
pseudodifferential operator with a classical symbol supported in a compact neighbourhood of
$p_0^{-1}(F)$. 
\subsection{The evolution operator} 

It is proved in \cite{ChaPo}, section 2, that there exists a Fourier Integral Operator (FIO)
$\tilde{U}_h$ which approaches modulo $O(h^{\infty})$, for $t\in \lbrack -T , T \rbrack^k,$ the operator
$$U_{\theta,h}(t)= e^{-ih^{-1}<t,P(h)>} \theta (P).$$ 
 
The general ``semi-classical FIO'' theory used by Charbonnel and Popov in \cite{ChaPo}
is based on the presentation of Duistermaat \cite{Du}.
(Remark that, in the context of first order classical elliptic operators on a compact manifold, Guillemin 
and Uribe have developed, in \cite{GU}, a FIO calculus for systems of commuting operators quite similar 
after the pionnier works of Colin de Verdi\`ere \cite{Cdv1} and \cite{Cdv2}).

\noindent More precisely $\tilde{U}_h\in I^{-k/4}(\R^{2n+k},\Lambda;h)$ with $C=\Lambda'$ the canonical 
relation
\begin{eqnarray}
C \ = \ \lbrace \  \Big((x,\xi),(y,\eta),(t,\tau)\Big) \in \ T^{ \ast } ({ \R }^{2n+k} )=
T^\ast\R^n\times T^\ast\R^n\times T^\ast\R^k
 \   ;    \nonumber \\ \tau = -
p_{0} (y , \eta ) \ , \ \ (x, \xi) = \Psi^t (y, \eta);\;(y, \eta ) \in { \cal O} 
    \   \rbrace
                                                         \label{eq:2.10}
\end{eqnarray}
where $\Psi^t$ is the joint flow of $p_{0}$ :
$$\Psi^t=\Psi_1^{t_1}\circ\dots\circ\Psi_k^{t_k}$$
and ${ \cal O}=q_0^{-1}(K)$ is a compact neighbourhood of $ \Sigma_0 $ which is invariant by the flow.

\noindent We parametrize $C$ by $(t,y, \eta ) \in \lbrack -T , T \rbrack^k \times { \cal O}$; 
then the principal symbol of $\tilde{U}_h$ is written :

\begin{equation}
\sigma(\tilde{U}_h)(t,y,\eta) = \exp\left(ih^{-1}(-\langle p_{0}(y,\eta), t\rangle + 
A(\gamma^t(y,\eta)))\right) \sigma_{1}\otimes\sigma_{2}
                                       \label{eq:symb}
\end{equation}
where $\sigma_1$ is the half density $|dt\wedge dy\wedge d\eta|^{1/2}$ and $\sigma_2$ is a ``fixed section 
of the Maslov bundle'', see below, $\gamma^t(y,\eta)$ is the  path
$$\gamma^t(y,\eta)=\{\Psi^{st}(y,\eta);0\leq s\leq 1\}
$$
and $A$ is the action of the path in $T^\ast(\R^n)$ :
$A(\gamma^t)=\int_{\gamma^t}\xi dx.$

\noindent Note that the function $\theta$ does not appear either in the Lagrangian manifold, nor in the 
principal symbol of $U_{\theta,h}(t)$, for $\theta (P)$ is a pseudodifferential operator, so its Lagrangian 
Manifold is the graph of the identity, and its principal symbol is equal to one in a small neighbourhood of 
$\Sigma_0$. 

\subsection{The action}
The manifold $\R^k.(y,\eta)=\{\Psi^t(y,\eta);\, t\in \R^k\}$ generated by the flow from one 
point is isotropic, because its tangent space admits the $K_j,\; 1\leq j\leq k$ as a basis ; on the other
hand $d(\xi dx)=\omega$ the symplectic form of $T^\ast(\R^n)$ ; we conclude then from the
Stokes formula that the action is constant on the homotopy class of a path in $\R^k.(y,\eta)$ and
$$A(\gamma^t(y,\eta))=\sum_{j=1}^{j=k} A(\gamma^{t_{j}}(\nu_j))$$ 
if $t=(t_1,\dots,t_k)$ ; $\nu_1=(y,\eta)$; 
$\nu_{j+1}=\Psi_j^{t_j}(\nu_j)$ and $\gamma^{t_{j}}(\nu_j)=\{\Psi_j^{s}(\nu_j);\, 0\leq s\leq t_j\}.$
But $\Sigma_0$ is a connected manifold and two points $\nu_0,\nu_1\in\Sigma_0$ can be connected
by a path $\nu_s.$ On $\{\Psi_j^{t_j}(\nu_s);\, t_j\in\R,0\leq s\leq 1\}$ the symplectic
form is null ( because $K_j\pdint\omega=0$ on $\Sigma_0$). Now if $T_j\in 2\pi\Z$ is a period the Stokes formula
gives that the action of the path $\gamma^{t_{j}}(\nu_s),\, t_j\in[0,T_j]$ does not depend on $s \in [0,1]$ and 
we have proved the
\begin{lemma}\ptir For all pair of points $\nu_0$ and $(y,\eta)$ in $\Sigma_0$ and for all periods
$T=(T_1,\dots,T_k)\in2\pi\Z^k$ we have
\begin{equation}
A(\gamma^T(y,\eta))=\sum_{j=1}^{j=k} A(\gamma^{T_{j}}(\nu_0)).
\label{eq:action}
\end{equation}
\end{lemma}

\subsection{The Maslov bundle}
We first recall the results of Arnol'd \cite{arnold}.
Let $\Ln$ be the Grassmannian manifold of the Lagrangian subspaces of $T^\ast\R^n$ 
and make the  identification $\Ln=U(n)/O(n).$ The application $Det^2$ is well define
on $\Ln.$ It is proved in \cite{arnold} that any path $\gamma :\ \Ss^1\to \Ln$ such that
$Det^2\circ\gamma\ :\,\Ss^1\to\Ss^1$ generates $\Pi_1(\Ss^1)$ gives a
generator of $\Pi_1(\Ln).$ Consequently $\Pi_1(\Ln)\simeq\Z$ and the cocycle $\mu_0$
defined by
$$\forall\gamma\in\Pi_1(\Ln)\quad\mu_0(\gamma)=\hbox{Degree}(Det^2\circ\gamma)$$
gives a generator of $H^1(\Ln)\simeq\Z.$
We can define a canonical {\em Maslov bundle} ${\mathbb M}(n)$ on $\Ln$ by the representation 
$\exp(i\frac{\pi}{2}\mu_0)=i^{\mu_0}$ of $\Pi_1(\Ln).$ This bundle is a bundle of torsion because
${\mathbb M}(n)^{\otimes 4}$ is trivial. 

Now the Maslov bundle of a Lagrangian submanifold ${\LL}$ 
of $T^\ast\R^n$ is the pull back of ${\mathbb M}(n)$ by the natural map
\begin{eqnarray*}
\varphi_n : \LL&\to& \Ln\\
\nu &\mapsto & T_\nu\LL.
\end{eqnarray*}
Arnol'd shows actually that $\mu={\varphi_n}^\ast\mu_0$ is the Maslov index of $\LL.$ It can be
written
\begin{eqnarray}
\mu : \Pi_1(\LL)&\to& \Z\nonumber\\
\lbrack\gamma\rbrack &\mapsto & <\mu_0,\varphi_n\circ\gamma>=\hbox{Degree}(Det^2\circ\varphi_n\circ\gamma).
\end{eqnarray}
We have to take care of the structural group of this bundle. As a $U(1)$
it is always trivial. But we will concider it at a $\Z_4=\{1,i,-1,-i\}$-bundle. Actually
we can see with the expression of the Maslov cocycle $\sigma_{jk}$ given in \cite{Ho2} (3.2.15) 
that the Maslov bundle has a trivial Chern class but $\sigma_{jk}$ can not be in general writen 
as a coboundary of a {\em constant} cochain. 

We recall now the result of the Proposition 3.2. p.132 of \cite{GS}.

\begin{prop}[Guillemin, Sternberg]\ptir Let $\Delta$ be an isotropic subspace of dimension
m in $T^\ast\R^{(n+m)}$ and define $S_\Delta=\{\lambda\in{\mathbb L}(n+m)/\;\lambda\supset\Delta\}.$ 
Then $S_\Delta$ is a submanifold of
${\mathbb L}(n+m)$ of codimension $(n+m)$, and if the map $\rho$ is defined by 
\begin{eqnarray*}{\mathbb L}(n+m)&\stackrel{\rho}{\to}&{\mathbb L}(n)\\
\lambda&\mapsto&\lambda\cap \Delta^\omega /\lambda\cap \Delta
\end{eqnarray*}
($\Delta^\omega$ denote the orthogonal of $\Delta$ for the canonical symplectic form 
$\omega$) then $\rho,$ which is not continue on ${\mathbb L}(n+m)$ itself, is smooth 
restricted to ${\mathbb L}(n+m)-S_\Delta$ and make this space as a fiber bundle
on ${\mathbb L}(n)$ with fiber $\R^{(n+m)}$. 

Moreover the image of the generator of 
$\Pi_1({\mathbb L}(n+m))$ is sent by $\rho$ to the generator of $\Pi_1({\mathbb L}(n))$. 
\end{prop}
This last result is easily seen if one choose symplectic coordinates $(x,\xi)$ such that $\Delta=\{x=0,\xi'=0\}.$ 
A generator of $\Pi_1({\mathbb L}(n+m))$ is given by $U_{(n+m)}(t)(\lambda_0)\,,\,0\leq t\leq 1$ where 
$\lambda_0=\{\xi=0\}$
and $U_{(n+m)}(t)$ is given in the complexe coordinates $z_j=x_j+i\xi_j$ by $U_{(n+m)}(t)(z_1,\dots,z_{(n+m)})=
(e^{i\pi t}z_1,z_2,\dots,z_{(n+m)}).$ The $(x',\xi')$ give symplectic coordinates of $T^\ast\R^{n}$
and $\rho(U_{(n+m)}(t)\lambda_0)=U_{n}(t)\lambda'_0.$

If we return now to $\Lambda$, we remark, as \cite{Ho4} p. 264 following \cite{DG} p.65, that 
\begin{lemma}the Maslov bundle of $\Lambda$ is trivial (as a $\Z_4$-bundle). 
\end{lemma}

{\sl proof}\ptir Actually $h_s(t,y,\eta)=(st,y,\eta)$ for
$0\leq s\leq 1$ makes a retrack $\Lambda_s$ of $\Lambda$ on $\Lambda_0$ such that
$\Lambda'_0=\{\Big( \nu,\nu,(0,-q_0(\nu)) \Big)\, ; \nu\in T^\ast\R^{n}\}$ and $\Lambda_0$ can be seen as a 
Lagrangian manifold in $T^\ast\R^{2n}.$ 
We now apply the proposition 2.1 with $\Delta=\Big\{\Big(0,0,(0,V)\Big)\in 
T^\ast\R^n\times T^\ast\R^n\times T^\ast\R^k\Big\}.$ The image $\varphi_{2n+k}(\Lambda_s)$ never meets
$S_\Delta$ so the Maslov bundle of $\Lambda$ is as well the pull-back by $\rho\circ\varphi_{2n+k}$
of ${\mathbb M}(2n)$, but there is an homotopy between $\rho\circ\varphi_{2n+k}(\Lambda)$ and
$\varphi_{2n}(\Lambda_0)$ and the application $\varphi_{2n}$ is constant on 
$\Lambda_0\subset T^\ast\R^{2n}.$\hfill\cqfd 

Then $\sigma_2=1$ in the formula of the principal symbol (\ref{eq:symb}).

\subsection{Composition}
Now we have to compose the FIO $\tilde{U}_h(t)$ with the operator $B(h)$ defined by :
\begin{eqnarray}
B(h) \ u \ (x) \ = \ {{1} \over {(2 \pi)}^k} \ \int e^{ih^{-1}<t, F>} \hat{\zeta}(t) \ u(x,t) \ dt          
\label{eq:B(h)}
\end{eqnarray}
$B(h)$ is a Fourier Integral Operator from $\R^k$ to $\R^0$ if we take
$x$ as a parameter. Its canonical relation  is equal to 
\begin{eqnarray}\label{eq:2.13}
C_B=\Lambda'_B \ = \ \lbrace \  (t,\tau) \in \ T^{ \ast } ({ \R^k }) \   ;  \tau = - F  \rbrace.
\end{eqnarray}          
We remark that the Maslov bundle of $C_B$ is trivial because the application $\varphi_k$ is
constant on it. Its principal symbol is equal to $ \sigma (t,-F) = {{1} \over {(2\pi)}^k} \ e^{ih^{-1}<t,F>} 
\hat{\zeta}(t) \vert dt \vert^{1/2}$ and $B(h)\in I^{-k/4}(\R^k,\Lambda_B;h)$.  
We have now to compose this two FIO. 
\begin{prop}  
We can approximate modulo $O(h^{\infty})$ the operator $\zeta
({{F-P} \over {h}}) \ \theta (P) \ $ by a FIO in $I^{-k/2}(\R^{2n},\Lambda_1;h)$ with 
\begin{eqnarray}
\Lambda'_1 =\ C\circ C_B= \ \lbrace \  (x,\xi,y,\eta) \in \ T^{ \ast } ({ \R }^{n}) \times T^{ \ast } ({ \R }^{n})
 \   ; 
 \exists t \in \R^k : (x,\xi) = \Psi^t (y,\eta) \nonumber, \  \\ 
    \  p_{0}(y, \eta ) = F 
\rbrace
          \label{eq:lagrange}
\end{eqnarray} 
and if we write for $(y,\eta)$ fixed, applying the hypothesis (H3),   
$${\lbrace t\,;\, (x,\xi)=\Psi^t(y,\eta) \rbrace}=t_0 + \oplus_{1\leq j\leq k}2 \pi \Z$$
the principal symbol is 
\begin{eqnarray}
\sigma_0(x,\xi,y,\eta) &=&  \sum_{\lbrace t\,;\, (x,\xi)=\Psi^t(y,\eta) \rbrace} 
\ {{1} \over
{(2 \pi)}^k}  \  (2 \pi h)^{-n/2} \ e^{ih^{-1}<t, F>} \hat{\zeta}(t) \times \nonumber \\
&&\quad\quad\quad\exp\left(ih^{-1}(-<p_{0} (y,\eta),t> +  A(\gamma^t))\right) \exp\left(i\frac{\pi}{2}\mu
(\gamma^t-\gamma^{t_0})\right)\ \sigma_1 \nonumber \\
&=&  \sum_{\lbrace t\,;\, (x,\xi)=\Psi^t(y,\eta) \rbrace} \ {{1} \over
{(2 \pi)}^k}  \  (2 \pi h)^{-n/2} \  \hat{\zeta}(t) \times \nonumber \\
&&\quad\quad\quad\quad\exp\left(ih^{-1}  A(\gamma^t)\right) \exp\left(i\frac{\pi}{2}\mu(\gamma^t-\gamma^{t_0})
\right)\ \sigma_1 .   \label{eq:symbA}
\end{eqnarray}
The sum in (\ref{eq:symbA}) is discrete and locally finite, because the support of $\hat{\zeta}$ is compact~;
$\sigma_1 $ is the canonical half density. 
\end{prop}
{\sl proof\ptir}
To understand the introduction of the Maslov term we have to 
make some recall on the composition of canonical relations. 

If $C_1$ is a canonical relation in $T^\ast \R^m\times T^\ast \R^n$ and $C_2$ 
a canonical relation in $T^\ast \R^n\times T^\ast \R^p$ the composition $C_1\circ C_2$
can be defined as a canonical relation in $T^\ast \R^m\times T^\ast \R^p$ if
$C_1\times C_2$ intersects transversally $T^\ast \R^m\times \Delta_{T^\ast \R^n}\times T^\ast \R^p$ 
(with the notation $\Delta_{T^\ast \R^n}=\{(Y,Y);\, Y\in T^\ast \R^n\}\subset({T^\ast \R^n})^2$). 
In fact it can be defined when the intersection is clean but in our case 
{\em this intersection is transversal} :\hfill\break
let $(\Psi^t(y,\eta),(y,\eta),(t,-F),(t,-F))$ be a point of this intersection. We will first calculate
the orthogonal, for the canonical scalar product, of the sum of the tangent spaces. 
If $(X,Y,T,T')\in T(T^\ast \R^n)\times T(T^\ast \R^n)\times T(T^\ast \R^k)\times T(T^\ast \R^k) $ is 
orthogonal to
$T(T^\ast \R^n\times T^\ast \R^n\times \Delta_{T^\ast \R^k})$ then it is of the form $(0,0,T,-T)$ now if it is
orthogonal to $(\sum u_j K_j, 0,(U,0),0)$ obtained by moving only the variable $t$ on $\Lambda$, and writing
$U=(u_1,\dots,u_k)$ then $T=(0,T_2)$, finally by moving only $(y,\eta)$ our vector must be orthogonal to
$(d\Psi^t(V),V,(0,-dp_0(V)),0)$ for any $V\in T_{(y,\eta)}T^\ast \R^n$ but $dp_0$ is surjective by hypothesis
(H1) so $T_2=0$. We will then follow \cite{Ho2}. 

If we denote by $C.C_B$ this intersection, then $C\circ C_B=\pi(C.C_B)$
where 
\begin{eqnarray*}\pi:T^\ast \R^{2n+k}\times {T^\ast \R^k}&\to &T^\ast \R^{2n}\\
(X,Y,T,T')&\mapsto &(X,Y)
\end{eqnarray*}
and 
$$\pi \, : C.C_B\to C\circ C_B $$
is a covering map. 

Indeed in our case 
\begin{eqnarray*} 
C.C_B&=&\left\{\left(\Psi^t(y,\eta),(y,\eta),(t,-F),(t,-F)\right);\;p_0(y,\eta)=F\right\}\\
C\circ C_B&=&\left\{\left((x,\xi),(y,\eta)\right);\;p_0(y,\eta)=F\hbox{ and }\exists\, t;\; (x,\xi)=
\Psi^t(y,\eta) \right\}
\end{eqnarray*}
So the fiber of $\pi$ is isomorphic to $2 \pi \Z^k$
(ie. $C\circ C_B=C.C_B/2 \pi \Z^k$) by hypothesis (H3), and $\pi$ realizes
an injection of $\Pi_1(C.C_B)$ in $\Pi_1(C\circ C_B)$ denoted by $\pi_\ast.$ We have then an exact 
sequence :
\begin{equation}
0\to\Pi_1(C.C_B)\stackrel{\pi_\ast}{\rightarrow}\Pi_1(C\circ C_B)\to 2 \pi \Z^k\to 0.
\label{eq:exact}\end{equation}
(See Theorem 3, Ch 4 § 19 in \cite{DNF}.)

Now if $M_{C},\, M_{C_B}$ and $M_{C\circ C_B}$ are the Maslov bundles of $C,\, C_B$
and $C\circ C_B$ respectively, one can make the following construction on $C.C_B$ :  
denoting by $\pi_1,\, \pi_2$
the projection of $C\times C_B$ on each factor, let 
$$M=\pi_1^\ast(M_{C})\otimes\pi_2^\ast(M_{C_B})_{|C. C_B}$$ 
be the restriction to $C.C_B$ of
$\pi_1^\ast(M_{C})\otimes\pi_2^\ast(M_{C_B})$, which is the Maslov bundle of $C\times C_B.$ 
By construction the bundle $M$ corresponds to the representation $i^\lambda$ 
of $\Pi_1(C.C_B)$ in $\C^\ast$ with :
$$\forall\gamma\in\Pi_1(C.C_B)\; ;\; \lambda(\gamma)=\mu(\pi_1\circ\gamma)+\mu(\pi_2\circ\gamma).
$$
But we have noted that $M_{C}$ and $M_{C_B}$ are trivial, it means that $\lambda(\gamma)=0$ in our case.
\begin{lemma}\ptir There is a natural relation  
$$M\simeq\pi^\ast M_{C\circ C_B}.$$ 
\end{lemma}
{\sl proof}\ptir We can consider this result as Theorem 21.6.7 in \cite{Ho3} ;
actually define $\Delta\subset T^\ast\R^{2n+2k}$
$$\Delta=\{ (0,0,V,V)\in T^\ast\R^{n}\times T^\ast\R^{n}\times T^\ast\R^{k}\times T^\ast\R^{k}\}$$
$\Delta$ is isotropic, one can identify $\Delta^\omega/\Delta\simeq T^\ast\R^{2n}$ as a 
symplectic space and we can apply the proposition 2.1. 
Consider the following diagram
\[\begin{array}{ccc}
C_1.C_2&\stackrel{\pi}{\longrightarrow}&C_1\circ C_2\\
\varphi_{2n+2k}\Big\downarrow\quad\quad&&\varphi_{2n}\Big\downarrow\quad\\
{\mathbb L}(2n+2k)&\stackrel{\rho}{\leadsto}&{\mathbb L}(2n)
\end{array}\]

where $\varphi_{2n+2k}(\nu)=T_\nu (C\times C_B).$
The map $\rho$ is defined as follow : 
$\rho(\lambda)=\lambda\cap \Delta^\omega /\lambda\cap \Delta$ for $\lambda\in{\mathbb L}(2n+2k).$ 
This map makes the diagram commutative. 

So the lemma is proved if one can see that the range of $\varphi_{2n+2k}$ is included
in \break${\mathbb L}(2n+2k)-S_\Delta$. Indeed for $\nu=(\Psi^t(y,\eta),(y,\eta),(t,-F),(t,-F))$
a vector in $T_\nu (C\times C_B)=T_{\pi_1(\nu)} C\times T_{\pi_2(\nu)} C_B$ is a sum of terms of
three types : $\Big(\sum \alpha_j K_j(\Psi^t(y,\eta)),0,(0,\alpha),(0,0)\Big)$ for $\alpha\in\R^k$,  
$\Big(d\Psi^t(Y),Y,(0,-dp_0(Y)),(0,0)\Big)$
for $Y\in T_{(y,\eta)}T^\ast\R^n$ and $\Big(0,0,0,(\beta,0)\Big)$ for $\beta\in\R^k$. But it is impossible
to write an element of $\Delta$ in such a way and we have $\varphi_{2n+2k}(\nu)\cap\Delta=\{0\}.$\hfill\cqfd

Consequence of the lemma :
\begin{equation}
\forall\gamma \in \Pi_1(C.C_B)\; ;\;\mu(\pi_\ast(\gamma))=\lambda(\gamma)=0.
\label{eq:nul}\end{equation}
Then all the non triviality of $M_{C\circ C_B}$ comes from the action of the Hamiltonian
flow, by (\ref{eq:exact}). It means, using the exact sequence (\ref{eq:exact}) that any section of 
$M_{C\circ C_B}$ can be represented by a $\C$-value function $f$ on $C.C_B$ which satisfies the 
equivariant relation :
\begin{equation}
\forall T\in 2 \pi \Z^k\ ;\quad f(t+T,y,\eta)=i^{-\mu(\gamma^T)}f(t,y,\eta)
\label{eq:true-equi}\end{equation}
if we parametrize $C.C_B$ by $(t,y,\eta)\in\R^k\times\Sigma_0$ and if we notice that, because
of the connexity of $\Sigma_0,$ the Maslov index of the loop 
$\gamma^T=\{(\Psi^{sT}(y,\eta),(y,\eta),0\leq s\leq 1\}$
is independent of the point $(y,\eta)$ ; because of the homotopy of the loops
$\gamma^T$ and $\gamma^{2\pi T_1 e_1}+\dots +\gamma^{2\pi T_k e_k}$ if 
$T=2\pi(T_1 e_1+\dots +T_k e_k)$ we can write :
\begin{equation}
\forall\  T\in \oplus_{1\leq j\leq k}2 \pi \Z\ ;\;
\mu(\gamma^T)=\sum_{j=1}^{j=k}T_j\mu_j\; \hbox{ with }\ \mu_j=\mu(\gamma^{2\pi e_j}).
\label{eq:maslov}
\end{equation}
Let now $\left((x,\xi),(y,\eta)\right)\in C\circ C_B.$ At each time that we choose $t\in\R^k$ such that
$(x,\xi)=\Psi^t(y,\eta)$ we have a natural local isomorphism between $M$ and $M_{C\circ C_B}$
as described in \cite{Ho2} p.181~; but when we change $t,$ say we take $t_0$ and $t_0+T$ with $T$
a period of our lattice, it corresponds to a change of trivialisation of  the bundle $M_{C\circ C_B}$
around our point. By definition of the Maslov bundle, the transition function is $i^{\mu(\gamma)}$
where $\gamma$ is the loop
$$\gamma (s)= \left(\Psi^{t_0+sT}(y,\eta),(y,\eta)\right)\qquad\gamma=\gamma^{t_0+T}-\gamma^{t_0}.$$
Notice that $\mu(\gamma)$ does not depend on the path that we draw in $C.C_B$ joining
the point $\left((x,\xi),(y,\eta),(t_0,-F),(t_0,-F)\right)$  to 
$\left((x,\xi),(y,\eta),(t_0+T,-F),(t_0+T,-F)\right)$
because of \break(\ref{eq:nul}).

The conclusion is that to define the product of the two symbols, we fix $t_0$ and then multiply
the product of the two symbols at $\left((x,\xi),(y,\eta),(t_0+T,-F),(t_0+T,-F)\right)$ by 
$i^{\mu(\gamma^{t_0+T}-\gamma^{t_0})}$ and make the sum for all periods $T$.
The result of this calculus is just (\ref{eq:symbA}) in the sens that the function
$\sigma_0(t_0,y,\eta)$ defined by the formula (\ref{eq:symbA}) is a $\C-$value function on $C.C_B$ 
witch satisfies the equivariance (\ref{eq:true-equi}).\hfill\cqfd

\noindent {\sl Remark}\ptir This lemma 2.3 is certainly the way to understand  that ``there is an isomorphism
between the two bundles'' (with our notations $M$ and $M_{C\circ C_B}$) as in \cite{Ho2} p.181 : in our 
case one is trivial but the second is not trivial in general so they cannot be really ``isomorphic''.
It would be compared also with Lemma 4.2 of \cite{BU} given without proof for the Maslov factor.

\subsection{end of the proof}
Let $T_0$ be a period of our lattice ; we want now to compare $\zeta ({{F-P(h)}\over{h}}) \theta (P)$ 
with\break
$\zeta ({{F-P(h)}\over{h}}) \exp-{{i}\over{h}}(<T_0,P(h)> )  \theta (P).$ One has 
\begin{eqnarray*}
\zeta ({{F-P(h)} \over {h}}) e^{-{{i}\over{h}}<T_0,P(h)>}\ \theta (P) \ &=& \ 
{{e^{-\frac{i}{h}<T_0,F>}} \over {(2 \pi)^k}} 
\int_{\R^k} e^{-\frac{i}{h}<t+T_0,(P(h)-F)>} 
\hat{\zeta}(t) \theta (P) dt \\
&=&\ {{e^{-\frac{i}{h}<T_0,F>} } \over {(2 \pi)^k}} 
\int_{\R^k} e^{-\frac{i}{h}<t,(P(h)-F)>} 
\hat{\zeta}(t-T_0) \theta (P) dt . 
\end{eqnarray*}
We can conclude by the previous calculus that $\zeta ({{F-P(h)}\over{h}})\exp-{{i}\over{h}}<T_0,P(h)>   \theta (P)$
can be approached by a FIO with the same Lagrangian than $\zeta ({{F-P(h)}\over{h}}) \theta (P)$
and with principal symbol 
\begin{eqnarray}
\sigma_{T_0}(x,\xi,y,\eta) &=& e^{-\frac{i}{h}<T_0,F>} \sum_{\lbrace t\,;\, (x,\xi)=\Psi^t(y,\eta) \rbrace} 
\ {{1} \over
{(2 \pi)}^k}  \  (2 \pi h)^{-n/2} \  \hat{\zeta}(t-T_0) \times \nonumber \\
&&\quad\quad\quad\exp\left(ih^{-1}  A(\gamma^t)\right) \exp\left(i\frac{\pi}{2}\mu(\gamma^t-\gamma^{t_0})
\right)
\ \sigma_1\nonumber \\
&=& e^{-\frac{i}{h}<T_0,F>} \sum_{\lbrace t\,;\, (x,\xi)=\Psi^t(y,\eta) \rbrace} 
\ {{1} \over
{(2 \pi)}^k}  \  (2 \pi h)^{-n/2} \  \hat{\zeta}(t) \times \nonumber \\
&&\quad\quad\quad\exp\left(ih^{-1}  A(\gamma^{t+T_0})\right) \exp\left(i\frac{\pi}{2}
\mu(\gamma^{t+T_0}-\gamma^{t_0})\right)\ \sigma_1 \\
&=&e^{i\left(h^{-1}  (A(\gamma^{T_0})-<T_0,F>)+\frac{\pi}{2}\mu(\gamma^{T_0})\right)} \sigma_{0}(x,\xi,y,\eta).
                                         \label{eq:symbA'}
\end{eqnarray}
We conclude that the two FIO $e^{-i\left(h^{-1}  (A(\gamma^{T_0})-<T_0,F>)+\frac{\pi}{2}\mu(\gamma^{T_0})\right)}
\zeta ({{F-P(h)}\over{h}}) e^{-{{i}\over{h}}(<T_0,P(h)>)}  \theta (P)$ and  $\zeta ({{F-P(h)}\over{h}}) \theta 
(P)$ have the same Lagrangian and the same principal symbol, if we 
remark that $A(\gamma^{T_0})$ and  $\mu(\gamma^{T_0})$ are 
numbers depending only on $T_0$ but not on $(y,\eta)\in \Sigma_0.$
So their difference is a FIO for which the amplitude has a compact support and a factor $h$~; we conclude by
the theorem of $L_2$ continuity (we can use the theorem of Asada-Fujiwara \cite{asa})
that there exists a constant $C>0$ such that
\begin{equation}\label{l2continu}
\|\zeta ({{F-P(h)}\over{h}})\left
({\rm Id}- e^{-{{i}\over{h}}(<T_0,P(h)-F>+A(\gamma^{T_0})+h\frac{\pi}{2}\mu(\gamma^{T_0}))}
\right)
\theta (P)\|_{L_2}\leq hC.
\end{equation}
{\sl Remark\ptir}We will give a precise estimate of the constant $C$ below with Lemma 3.2.

Now if $f$ is a commun eigenfunction of our $P_j(h)$ with joint eigenvalue 
$\lambda=(\lambda_1,\dots,\lambda_k)$
and if we suppose $\|F-\lambda\|\leq ch$ and $|\zeta|\geq d>0$ on the set 
$\{x\in\R^k;\, \|x\|\leq c\},$ the previous evaluation gives
\[|1-e^{-{{i}\over{h}}(<T_0,\lambda-F>+A(\gamma^{T_0})+h\frac{\pi}{2}\mu(\gamma^{T_0}))}|\leq h
\frac{C}{d}.\]
This gives the theorem by taking for $T_0$ the basis of the lattice.\hfill\cqfd

\subsection{extension}

What happens if one relaxes the hypothesis ($H_3$) ? Following the formula
(2.11) of \cite{ChaPo} we see that appears a new term in the principal symbol of the evolution
operator :
$$\exp\left(-i\int_0^1<p_1(\Psi^{st}(y,\eta)),t>ds\right)$$
and in the comparison between the principal symbol of $\zeta ({{F-P(h)}\over{h}}) e^{-{{i}\over{h}}
(<T_0,P(h)>)}  \theta (P)$ and  $\zeta ({{F-P(h)}\over{h}})\ \theta (P)$ will appear a new term which
a priori depends on $(y,\eta)$ :
$$\exp\left(-i\int_0^1<p_1(\Psi^{sT_0}(y,\eta)),t>ds\right).$$
The hypothesis ($H'_3$) assures that this term is constant, let us denote it by $e^{-i\delta(T_0)}$. 
Following the preceding proof we obtain
\begin{theo}\ptir
Under the assumptions $H_1$, $H_2$, $H'_3$ and $H_4$, the part of the joint spectrum 
$\Lambda ^{{\it Q} (h)}$ 
lying in any
$k$-cube $ \displaystyle { \prod_{j=1}^{k}} \ \rbrack E_{0j} -h c_j \ , \ E_{0j} + h c_j \lbrack$  centered at $E_0$ 
is localized
modulo $O(h^2)$ near a lattice 
$$E_0 \ + \  {\bf a}^{-1} 
\Big( ((\frac{\delta_{1}}{2\pi}-\frac{ \mu_1}{4}) h - \frac{\alpha_1}{2\pi} + \Z h)
\oplus
\ldots \oplus ((\frac{\delta_{k}}{2\pi} - \frac{ \mu_k}{4}) h - \frac{\alpha_k}{2\pi} + \Z h) \Big), $$ where 
$\delta_{j}$ are the 
integral of the subprincipal symbol on the basic cycles of the torus acting on $\Sigma_0$,
the  $\mu_j$ are the Maslov indices of these cycles and $\alpha_j$ are the action of these cycles. 
\end{theo}

\section{Multiplicity}
For simplicity we maintain  in this section the hypothesis ($H_3$).

\begin{theo}\ptir For $\nn\in\Z^k$ denote by $I_\nn(h)$ the $k$-cube with size $2Ch^2$ centered at
$E_{0}+\A^{-1}(-\frac{\alpha}{2\pi}-\frac{\mu}{4}h+\nn h)$ 
and suppose that $I_\nn(h)\subset\displaystyle { \prod_{j=1}^{k}}]E_{0j}-c_j h,E_{0j}+c_j h[.$ Then the number
\[N_\nn(h)=\sharp \Big(I_\nn(h)\cap{\rm Spec}(P(h))\Big)\] admits the following behaviour on $h$ :
there exists a sequence $\Big(l_j(\nn)\Big)_{j\in\N}$ such that for any $m\in \N$
\begin{equation}N_\nn(h)=h^{k-n}\sum_0^{m-1}l_j(\nn)h^j+O(h^{m-n})\label{dev}
\end{equation}
and $l_0(\nn)=l_0=\frac{1}{(2\pi)^n}\int_{\Sigma_0} d\nu$ where $d\nu$ denote the Liouville density of
$\Sigma_0.$
\end{theo}
As before, we work with the new operators $P(h).$
The Liouville density of $\Sigma_0$ is defined as follows : the map $p_0 : T^\ast\R^n\to\R^k$
is a submersion on a \nbd of $\Sigma_0$ because of hypothesis ($H_1$) and then defines a density
on $\Sigma_0$ by ``dividing'' the euclidean density $dx_{T^\ast\R^n}$ by the pullback 
$p_0^\ast(dx_{\R^k})$ of the euclidean density of $\R^k$ by the submersion. Actually 
$p_0^\ast(dx_{\R^k})$ is a well defined density on a transversal of $\Sigma_0$ and $d\nu$ must
satisfy at the points of $\Sigma_0$
\[dx_{T^\ast\R^n}=d\nu\wedge p_0^\ast(dx_{\R^k}).\]
If we use the Riemannian structure of the submanifold $\Sigma_0$ of the euclidean space $T^\ast\R^n$
and denote by $dS$ the associated Riemannian density, then
$$dS=\|dp_{01}\wedge\dots\wedge dp_{0k}\|d\nu.$$ 
{\sl proof}\ptir We will just sketch it because we just follow \cite{kn:D} herself inspired by \cite{BU}.
We want to approach $N_\nn(h)$ by a trace 
\[{\rm Tr}_\zeta(h)={\rm Tr}\Big(\zeta ({{F-P(h)}\over{h}}) \ \theta (P(h))\Big) = 
{\rm Tr}\Big( {{1}\over{(2 \pi)^k}} \int_{\R^k} e^{-\frac{i}{h}<t,(P(h)-F)>} \hat{\zeta}(t)\theta (P(h)) dt
\Big) \]
with $\zeta$ to be chosen satisfying two conditions : $\hat{\zeta}(0)=1$ and the support of $\hat{\zeta}$ 
is a small compact such that 0 is the only one period of the joint flow belonging to Supp$(\hat{\zeta}).$

On one hand we can calculate ${\rm Tr}_\zeta(h)$ using the stationnary phase theorem and obtain the 
development (\ref{dev}) with the expression of $l_0$ as mentioned in \cite{ChaPo} Theorem 5.2.

On the other hand we cut ${\rm Tr}_\zeta(h)$ in three terms
\[{\rm Tr}_\zeta(h)=\Big(\sum_{\lambda ; \ \forall j |\lambda_j-F_j|<hc_j }+
\sum_{\stackrel{\lambda ;\ \forall j  |\lambda_j-F_j|<h^{1-\epsilon}c_j}{\exists j |\lambda_j-F_j|\geq hc_j}}
+\sum_{\lambda ;\exists j |\lambda_j-F_j|\geq h^{1-\epsilon}c_j}\Big)
\zeta({F-\lambda\over h})\theta(\lambda)\]
and we will choose $\epsilon$ and $\zeta.$ The third term can be controlled by $O_\zeta(h^{\epsilon N})$ 
for any $N$ because $\zeta\in{\cal S}(\R^k).$ We just have to take $N$ such that $\epsilon N>m+k-n$ when 
$\epsilon$ is fixed.
If we want that the first term approaches $N_\nn(h)$ we must choose $\zeta$.
\begin{lemma}\ptir For any $N\in\N$ there exists a function $\varphi_N:\R^k\to\R$ satisfying
\begin{eqnarray}
\hat\varphi_N\in C^\infty_0(\R^k),\quad\varphi_N(t)\stackrel{0}{\simeq}1+O(\|t\|^N)\hbox{ and}\label{bif}\\
\exists c,\,\forall \,\mm\in\Z^k-\{0\}\quad|\varphi_N(t)|\leq c\|t- \mm\|^N\label{bof}.
\end{eqnarray}
\end{lemma}
{\sl proof}\ptir Remark first that for a function $f,\hat f\in C^\infty_0(\R)$ and $f(0)=1,$ the function
$\varphi(x)=f(x)\frac{\sin(2\pi x)}{2\pi x}$ satisfies 
\[\hat\varphi\in  C^\infty_0(\R)\, ;\quad \varphi(x)\stackrel{0}{\simeq}1+O(|x|)\, ;\quad\exists c>0\, ; 
\forall n\in\Z-\{0\},\,|\varphi(x)|\leq c |x-n|
\]
(see Lemma 2.5.1 of \cite{kn:D}). The function 
\[\varphi_N(t_1,\dots,t_k)=\prod_{j=1}^k \Big(1-\big(1-\varphi(t_j)\big)^N\Big)^N
\]
satisfies the properties of the lemma.\hfill\cqfd

Define now for any $\mm\in\Z^k$
\begin{equation}\label{zeta}
\beta_\mm=\frac{1}{2\pi h}\alpha+\frac{1}{4}\mu -\mm,\quad\hbox{and}\quad\zeta(t)=\varphi_N(t-\beta_\nn)
\end{equation}
then, by the Theorem 1 
\begin{eqnarray*}
\sum_{\lambda ; \ \forall j |\lambda_j-F_j|<hc_j }\zeta({F-\lambda\over h})\theta(\lambda)&=&
\sum_{\mm\in\Z^k}\left(\sum_{\lambda ;\|\lambda-(F-h\beta_\mm)\|<Ch^2}\zeta({F-\lambda\over h})\theta(\lambda)\right)\\
&=&\sum_{\lambda ;\|\lambda-(F-h\beta_\nn)\|<Ch^2}\zeta({F-\lambda\over h})\theta(\lambda)+O(h^{N+k-n})
\end{eqnarray*}
if we use (\ref{bof}) to bound each other term and Lemma 5.4 of \cite{ChaPo} to bound the number of such terms 
by $O(h^{k-n})$. Then using (\ref{bif}) we obtain
\begin{cor}\ptir For any $m\in\N$ if we define $\zeta$ by the formula above with $N>m$ then
\[|\ N_\nn(h)-\sum_{\lambda ; \ \forall j |\lambda_j-F_j|<hc_j }\zeta({F-\lambda\over h})\theta(\lambda)\ |\leq Ch^{m+k-n}
\]
\end{cor}
For the second term we need a refined version of Theorem 1 by the control of the constant C
in the inequality (\ref{l2continu}).
\begin{lemma}\ptir For a function $\zeta$ such that $\hat\zeta\in C_0^\infty(\R^k)$ and $b\in\R^{>0}$
define $\zeta_b(t)=\frac{1}{b^k}\zeta(\frac{t}{b}).$ There exists $b_0,h_0,C>0$ such that for any
$h<h_0$ and $b>b_0,\,$
\[
\|\zeta_b ({{F-P(h)}\over{h}})\left
({\rm Id}- e^{-{{i}\over{h}}(<T_0,P(h)-F>+A(\gamma^{T_0})+h\frac{\pi}{2}\mu(\gamma^{T_0}))}
\right)
\theta (P)\|_{L_2}\leq b^{2-k}hC.
\]
\end{lemma}
{\sl proof}\ptir We will write here the kernel of the evolution operator as an oscillatory integral.
If we remark that $\hat{\zeta_b}(t)=\hat{\zeta}(bt),$ the kernel of the operator 
\hfill\break$\zeta_b ({{F-P(h)}\over{h}})\left
({\rm Id}- e^{-{{i}\over{h}}(<T_0,P(h)-F>+A(\gamma^{T_0})+h\frac{\pi}{2}\mu(\gamma^{T_0}))}
\right)\theta (P)$ can be written (locally in $(x,y)$) 
\[\int\hat\zeta(bt)e^{\frac{i}{h}(\phi(t,x,y,\theta)+<t,F>)}a(t,x,y,\theta,h)dt\ d\theta=
\int e^{\frac{i}{h}\tilde{\phi}(t,x,y,\theta)}\tilde{a}(t,x,y,\theta,h)dt\ d\theta.
\]
The relation between the phase function and the amplitude with the geometric objects already described are :
$C=\{(x,y,t,\phi'_x,\phi'_y,\phi'_t);\phi'_\theta(t,x,y,\theta)=0\}.$  
The amplitude $\tilde{a}$ has compact support and by the previous section
we know that $\tilde{a}(t,x,y,\theta,h)=h^{-n}\sum h^j\tilde{a}_j(t,x,y,\theta)$ and
$\tilde{a}_0(t,x,y,\theta)$ vanishes on the canonical relation of our operator~: $\Lambda'_1.$ Moreover for 
any integer $l$ there exists a constant $C_l>0$ such that 
\begin{equation}\label{born}
\|D^l\tilde{a}_0\|_\infty\leq b^l C_l
\end{equation}
where $D^l$ is any composition of $l$ 
partial derivatives. Finally we can remark that, the phase function $\tilde{\phi}$ is non-degenerated in the 
sense that the function $(x,y,t,\theta)\to(\phi'_t+F,\phi'_\theta)$ is a submersion. This fact is a 
consequence of the description of $\phi$ given in \cite{ChaPo} Theorem 4.2.

\noindent Because the phase function is non-degenerated and $\tilde{a}_0$ vanishes on $\Lambda'_1$ which
can be described as $\Lambda'_1=\{(x,y,\tilde{\phi}'_x,\tilde{\phi}'_y);\tilde{\phi}'_t(t,x,y,\theta)=0,
\tilde{\phi}'_\theta(t,x,y,\theta)=0\},$ there exist $C^\infty$-functions with compact support
$\gamma_j$ and $\chi_l$ such that $\tilde{a}_0=<\gamma,\tilde{\phi}'_t>+<\chi,\tilde{\phi}'_\theta>,$
and as they are defined in terms of derivatives of $\tilde{a}_0$ and as $\tilde{a}_0$ satisfies
(\ref{born}) their derivatives of order $l$ are controlled by $b^{l+1}.$
Finally an integration by parts gives
\begin{eqnarray*}
\int e^{\frac{i}{h}\tilde{\phi}(t,x,y,\theta)}
\left(<\gamma,\tilde{\phi}'_t>+<\chi,\tilde{\phi}'_\theta>\right)&=&\\
\frac{-h}{i}\int e^{\frac{i}{h}\tilde{\phi}(t,x,y,\theta)}\Big(\sum_j\frac{\partial\gamma_j}
{\partial t_j}+\sum_l\frac{\partial\chi_l}{\partial \theta_l}\Big)&=&O(hb^{2-k})
\end{eqnarray*}
if we remember that Supp$(\hat{\zeta_b})\subset]\frac{-1}{b},\frac{1}{b}[^k.$\hfill\cqfd
\begin{cor}\ptir For any $\epsilon<\!\!<1/2$ if $\lambda=\lambda(h)$ is a joint eigenvalue of $P(h)$ such 
that\break$\|\lambda-F\|_\infty\leq ch^{1-\epsilon}$ then there exist $C>0$ and $\mm\in\Z^k$ such that
\[\|\lambda-(F-h\beta_\mm)\|\leq Ch^{2-2\epsilon}
\]
with $\beta_\mm$ defined by (\ref{zeta}).
\end{cor}
{\sl proof}\ptir We use Lemma 3.2 with $b=h^{-\epsilon}$ remarking that $|\zeta_b(t)|\geq \frac{d}{b^k}$
if $\|t\|_\infty \leq bc.$ It follows that if $\|\lambda-F\|_\infty\leq ch^{1-\epsilon}$ then
\[|1-e^{-{{i}\over{h}}(<T_0,\lambda-F>+A(\gamma^{T_0})+h\frac{\pi}{2}\mu(\gamma^{T_0}))}|\leq h^{1-2\epsilon}
\frac{C}{d}.
\]
The conclusion goes as for Theorem 1.\hfill\cqfd

\noindent {\sl Consequence}\ptir If $\lambda$ is a joint eigenvalue occurring in the second term of ${\rm Tr}_\zeta(h)$
for the special $\zeta$ defined by (\ref{zeta}), then it can be written $F-h\beta_\mm+O(h^{2-2\epsilon})$ 
with $\mm\neq\nn$ an  $\zeta(({F-\lambda\over h})\theta(\lambda)=O(h^{N(1-2\epsilon)})$ by (\ref{bif}) ; 
as previously we can bound the number of terms occurring in this sum by $O(h^{k-n}),$ we obtain finally
\[|\sum_{\stackrel{\lambda ;\ \forall j  |\lambda_j-F_j|<h^{1-\epsilon}c_j}{\exists j |\lambda_j-F_j|\geq hc_j}}
\zeta({F-\lambda\over h})\theta(\lambda)\ |\leq Ch^{N(1-2\epsilon)+k-n}
\]
it can be written as $O(h^{m+k-n})$ if $N(1-2\epsilon)>m.$ Therefore we have proved that for any integer $m$
there exist $C>0$ and a function $\zeta$ such that $\hat\zeta(0)=1$ and $0$ is the only period of the joint flow
belonging to the (compact) support of $\hat\zeta$ and such that
\[|\ {\rm Tr}_\zeta(h)-N_\nn(h)\ |\leq Ch^{m+k-n}
\]
for $h$ small enough.
This concludes the proof of Theorem 3.\hfill\cqfd

\end{document}